\documentclass{bmvc2k}

\title{AGMDT: Virtual Staining of Renal Histology Images with Adjacency-Guided Multi-Domain Transfer}

\addauthor{Tao Ma}{matao@stu.pku.edu.cn}{1*}
\addauthor{Chao Zhang}{chaozhang@stu.pku.edu.cn}{1*}
\addauthor{Min Lu}{lumin@bjmu.edu.cn}{2**}
\addauthor{Lin Luo}{luol@pku.edu.cn}{1**}

\addinstitution{
 College of Engineering\\
 Peking University\\
 Beijing, China
}
\addinstitution{
Department of Pathology\\
School of Basic Medical Sciences\\
Peking University Third Hospital\\
Peking University Health Science Center\\
Beijing, China
}

\runninghead{Tao Ma, Chao Zhang, Min Lu, Lin Luo}{AGMDT}

\def\etal{\emph{et al}\bmvaOneDot}

\usepackage{amssymb}

\begin{document}

\maketitle

\begin{abstract}
Renal pathology, as the gold standard of kidney disease diagnosis, requires doctors to analyze a series of tissue slices stained by H\&E staining and special staining like Masson, PASM, and PAS, respectively. These special staining methods are costly, time-consuming, and hard to standardize for wide use especially in primary hospitals. 
Advances of supervised learning methods have enabled the virtually conversion of H\&E images into special staining images, but achieving pixel-to-pixel alignment for training remains challenging.
In contrast, unsupervised learning methods regarding different stains as different style transfer domains can utilize unpaired data, but they ignore the spatial inter-domain correlations and thus decrease the trustworthiness of structural details for diagnosis. In this paper, we propose a novel virtual staining framework AGMDT to translate images into other domains by avoiding pixel-level alignment and meanwhile utilizing the correlations among adjacent tissue slices. We first build a high-quality multi-domain renal histological dataset where each specimen case comprises a series of slices stained in various ways. Based on it, the proposed framework AGMDT discovers patch-level aligned pairs across the serial slices of multi-domains through glomerulus detection and bipartite graph matching, and utilizes such correlations to supervise the end-to-end model for multi-domain staining transformation. 
Experimental results show that the proposed AGMDT achieves a good balance between the precise pixel-level alignment and unpaired domain transfer by exploiting correlations across multi-domain serial pathological slices, and outperforms the state-of-the-art methods in both quantitative measure and morphological details.  
\end{abstract}

\section{Introduction}
\label{sec:intro}
Pathological examination is the gold standard of clinical diagnosis. One essential step is the staining in pathological slide preparation, which aims to highlight tissue structures and enhance lesion visibility. In renal pathology, pathologists utilize four types of stains, namely: H\&E, Masson, PASM, and PAS, each representing unique structural features for diagnosis. For example, the basic H\&E staining presents cell nucleus to be blue and cytoplasms to be pink, while PAS stains the contours of extracellular matrices such as the basal membrane of glomerulus (kidney spherule), tubules, mesangial matrix to facilitate identification of inherent cell types. Compared to PAS, PASM performs better for thickened basal membrane lesions and stains the basal membrane, mesangial matrix, and type IV collagen black to identify cell types by their location to the basal membrane of glomerulus. Masson's trichrome stain is widely used for staining collagen fibers, where it portrays the basal membrane and type III collagen as blue or green, and immune complexes, plasma, and fibrinogen as red. Among the four types, H\&E staining is the most commonly used, while the three special staining methods also have their unique value in renal pathological diagnose.
However, since more samples need to be taken, and the procedures of special staining are complex with the staining effects highly relying on experienced operators, multi-staining increases the cost and uncertainty of diagnosis significantly. 

In the field of histopathology-assisted diagnosis, some deep learning-based methods\cite{Qin_2022_ACCV} have proven effective. The advances of deep learning technologies also trigger virtual staining as a promising alternative. Leveraging Generative Adversarial Network (GAN)\cite{NIPS2014_5ca3e9b1}, researchers can transform stained tissues from one type to another digitally. Rivension \etal \cite{rivenson2019virtual} uses deep learning on autofluorescence images of unlabeled histology samples to convert unstained slices into images of various staining effects, where the virtually-stained images closely match the ones under standard chemical staining. On this basis, de Haan \etal \cite{de2021deep} employes autofluorescence as an intermediate modality and trained a supervised neural network to virtually stain H\&E histological images into virtually-stained images of special stains such as Masson, PAS, and silver staining (PASM). 
Within deep learning technologies, supervised methods exhibit high accuracy and reliability, but since one tissue slice cannot be stained multiple times, pixel-to-pixel cross-domain alignment is hard to achieve, and additional channels such as autofluorescence must be used as mediators. Zeng \etal \cite{zeng2022semi} proposed a semi-supervised virtual staining method to associate adjacent slices utilizing patch labels of binary progesterone receptors (PR) results and transfer H\&E images into immunohistochemistry (IHC) images. But though it can preserve structural consistency between H\&E and IHC slices, it highly depends on the binary PR labels and is not easy to generalize into other staining types.

There are also some unsupervised methods applied to staining transfer. UGATIT \cite{kim2019u} is proven to visually perform well in single-modal unsupervised stain translation task. Further, based on StarGAN\cite{choi2018stargan}, Lin \etal \cite{lin2022unpaired} proposed UMDST, a multi-domain stain transfer method for kidney histological images, using a single network to generate multiple types of virtually stained images.
However, due to the ignorance of spacial alignment of structural details, unsupervised methods sometimes tend to produce pseudo staining artifacts of anatomical structures, especially for important diagnostic features such as glomerular structures.


This paper proposes a novel adjacency-guided multi-domain transfer framework (AGMDT) to transfer renal histology images into multiple staining types, exploiting the correlations across adjacent slices with different stains. 
The framework uses a generator-discriminator basis, where an adaptive paring module is introduced using glomerulus detection and bipartite graph matching to obtain patch-level aligned pairs in adjacent slices, and a multi-domain stain transferring module guides the training process with the correlations across pairs from adjacent tissue slices. Special loss function is designed to reflect the patch-level adjacent-slice correlations when adjusting the network. 
We also build a full-stack renal histological dataset. Each case in our dataset has all four staining types: H\&E, Masson, PASM, PAS that applied to adjacent slices of the same tissue. In addition, the dataset contains 32,413 pairs of glomeruli aligned at the patch level.

To our knowledge, this is the first uniform virtual staining framework that can generate multi-stained histological images utilizing correlations across adjacent slices. In addition, we provide a high-quality renal histological dataset for future researchers to train and validate virtual staining methods. Experimental results show that the proposed AGMDT framework outperforms the state-of-the-art staining methods in quantitative measures and meanwhile represents more realistic and accurate structural details for better clinical diagnosis.

\begin{figure*}[t]
\centerline{\includegraphics[width = \textwidth]{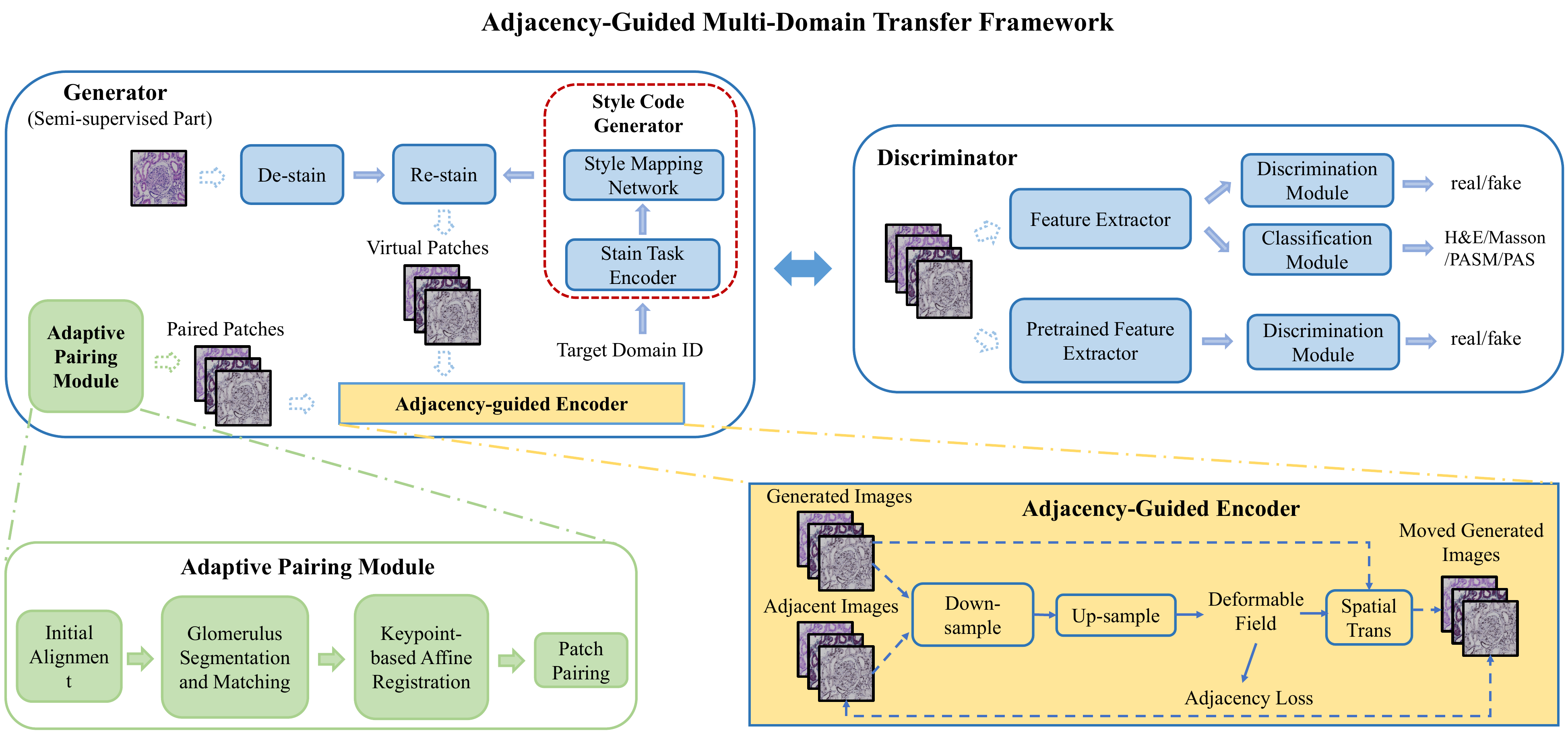}}
\caption{An overview of the adjacency-guided multi-domain transfer framework.} \label{overview}
\end{figure*}

\section{Related work}

\paragraph{Medical Image-to-image Translation} I2I translation involves transforming an image from one style to another while preserving its content. In the field of biological imaging, image translation techniques are used for tasks such as color normalization, virtual staining, and MRI modality conversion\cite{lin2022unpaired}. Compared to natural image translation tasks, image translation tasks in the area of biological imaging require higher accuracy and correctness in the generated images. However, in many cases, obtaining pixel-level paired data needed for supervised methods is extremely difficult. On the other hand, unsupervised methods are prone to producing artifacts that fail to meet the requirements. To address these challenges, some researchers have attempted to leverage adjacency information and domain knowledge to perform semi-supervised learning using non-pixel-accurate aligned paired data\cite{kong2021breaking, zeng2022semi}, but these methods do not readily apply to the task of I2I translation for special stains. 

\paragraph{Histological Image Registration} Medical image registration is an essential step in many computer-aided medical image analysis tasks\cite{ge2022unsupervised}. Histological image registration is particularly challenging due to the large sizes of some images and differences in local structure between slices\cite{borovec2020anhir}. Common image registration approaches such as intensity-based and feature-based methods that use hand-crafted image features\cite{sotiras2013deformable} cannot be applied directly. Song \etal\cite{song2013unsupervised} developed an unsupervised content classification method that generates multichannel probability images from a roughly aligned image pair, enhancing the structural similarity between image pairs. The emergence of the ANHIR challenge has given new vitality to this field. The lead method proposed by the Mevis team\cite{lotz2019robust} is a 3-step registration pipeline consisting of  robust pre-alignment, NGF similarity-based iterative affine registration and B-spline-based non-rigid registration, which works well but may introduce additional overhead. Existing methods have some limitations in the registration of adjacent slices in virtual staining of renal histology images. The registration process does not include features unique to renal histology images like glomerular morphology and location and may be time-consuming due to the large size of histology images.

\section{Method}



Inspired by pathologists' diagnostic work with adjacent slices of different staining, we propose the AGMDT method for multi-domain stain transfer that incorporates information from adjacent slices. We introduce an adaptive pairing module and an adjacency-guided encoder module to generate and incorporate supervision information from these adjacent slices and better guide the process of stain transfer. In this way, we avoid the limitation of pixel-level aligned data that impede the supervised methods, and achieve better transformation results than unsupervised methods.

\subsection{Adjacency-Guided Multi-Domain Transfer Framework}

Figure 1 shows the overall structure of AGMDT which is a GAN-based multi-domain stain transfer framework. Our generator comprises three key components: a basic encoder-decoder structure for de-stain and re-stain, a style code generator, and the adjacent supervision component. Specifically, the encoder extracts features from the input image, while the decoder reconstructs the image with the guidance of the target domain style code generated by the style code generator and the input image features extracted by the encoder. To better represent the target domain features, we enhance the style code generator by using a 64-dimensional vector as the stain domain label and by utilizing a multi-head attention layer to extract task-related features. 

The adjacent supervision component is the key component which consists of two modules: the adaptive pairing module and the adjacency-guided encoder module. Among them, the adaptive pairing module provides paired patches of the multi-domain renal histological dataset to the adjacency-guided encoder module, followed by stain transfer supervision with the aid of the adjacency-guided encoder module. The structure of the adjacency-guided encoder module is based on U-net\cite{ronneberger2015u}. It takes a pair of adjacent-slice patches and the corresponding generated patch as input. By predicting the deformable field, we can obtain the moved generated image using this field. Furthermore, the distance between the adjacent-slice and moved generated images is calculated to encourage the model to learn a more realistic style.

The discriminator component of AGMDT adopts a dual-discriminator structure: one of the discriminators has the same structure as the discriminator in UMDST and another employs a pre-trained large model structure, which utilizes a frozen pre-trained backbone to extract image features and a learnable head to judge whether an image is real or generated \cite{kumari2022ensembling}. The dual-discriminator structure enables the model to effectively distinguish between generated and real images, and to guide the generator in performing virtual staining.

\subsection{Adaptive Pairing Module}
To maximize data utilization, we proposed the adaptive pairing module, which adaptively provides paired patches of the multi-domain renal histological dataset to the semi-supervised part of the model. As shown in Figure 2, the adaptive pairing module completes its task in four stages: initial alignment, glomerulus segmentation and matching, keypoint-based affine registration and patch pairing. 

The goal of initial alignment is to achieve a rough but fast contour alignment. Due to the inherent structural differences in adjacent tissue slices, only one parameter, the rotation angle, needs to be optimized in this stage\cite{wodzinski2021deephistreg}. After calculating the center point of the source and target images, the adaptive pairing module used an exhaustive rotation angle search method to obtain the initial alignment result.

In the glomerulus segmentation and matching stage, it is imperative to accomplish the detection and matching of glomeruli in adjacent slice pairs and to store the center coordinates of the paired glomeruli as keypoint information. First, the module used a pre-trained model\cite{jiang2021deep} to segment glomeruli on adjacent slice pairs and record their location in the original image. Then the relationship between the center distances of the glomeruli was converted into a bipartite graph matching problem, which was then solved using the Hungarian algorithm to match the glomeruli. Finally, the center coordinates of the successfully matched glomerulus image pairs are stored as keypoint pairs.

Since there are still a large number of kidney tissue areas that cannot be registered, affine registration based on keypoints\cite{marstal2016simpleelastix} was performed to improve the overall registration effect of the kidney tissue. After the first three stages, we obtained a region-level aligned dataset of whole slide images (WSI) along with keypoint pairs of the glomeruli. The region-level aligned dataset is divided into groups, and each group contains H\&E/PAS/PASM/Masson-stained adjacent slices of kidney tissue. 

In the patch pairing stage, the adaptive pairing module combined images of H\&E and the other three staining types in each group into image pairs, and cut patches by pairs. After calculating the similarity of H\&E-PAS/PASM/Masson stained patch pairs using lpips\cite{zhang2018unreasonable} and fsim\cite{zhang2011fsim}, it filtered out the successfully paired patches according to the similarity, which were fed into the semi-supervised part of the generator. Meanwhile, the unpaired patches were fed into the unsupervised part of the generator for training, making full use of the data in this way. In the process, the module also reused the key point pairs to build a glomerulus-aligned renal histological dataset.

\begin{figure*}[htbp]
\centerline{\includegraphics[width = \textwidth]{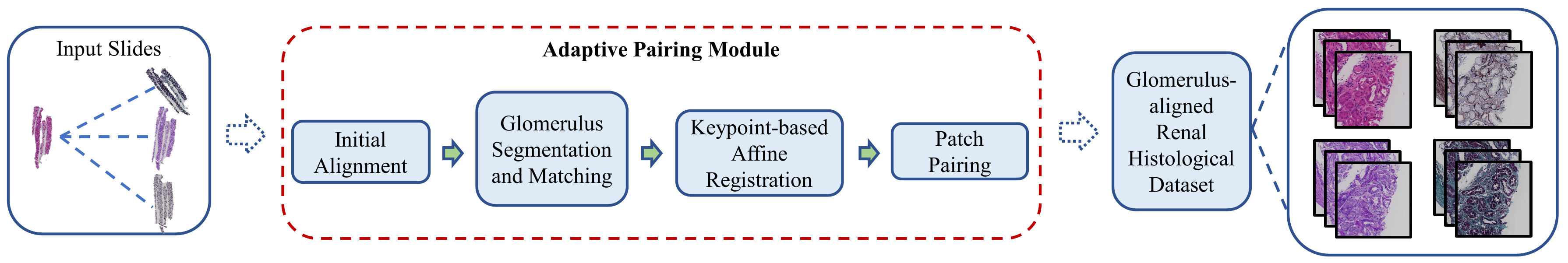}}
\caption{The structure of adaptive pairing module, including initial alignment, glomerulus segmentation and matching, keypoint-based affine registration and patch pairing. } \label{Registration}
\end{figure*}

\subsection{Loss Function}
To incorporate adjacent-slice information as supervision, we design the following loss function. The generator's loss function consists of six components, including: adversarial loss$l^{G}_{adv}$, classification $l^{G}_{cls}$, auxiliary loss $l^{G}_{\eta adv}$, $l^{G}_{\eta cls}$,cycle loss $l_{cyc}$, identity loss $l_{idt}$, and adjacency-guided loss  $l_{adj}$, which measures the discrepancy between the generated data and its corresponding adjacency data. The loss function is formulated as follows:
\begin{equation}
\begin{split}
\begin{aligned}
l^{G} = \lambda_{1} \times (l^{G}_{adv} + l^{G}_{cls} + l_{cyc} + l_{idt})+\lambda_{2} \times(
l^{G}_{\eta adv} + l^{G}_{\eta cls}) + \lambda_{3} \times l_{adj}
\end{aligned}
\end{split}
\end{equation}

We denote the Adjacency-guided Encoder as \textit{A}, the encoder-decoder module for de-stain and re-stain as \textit{S} and the formula for computing $l_{adj}$ is as follows:
\begin{equation}
    l_{adj}=\mathbb{E}_{x, \widetilde{y}}\left[\|\nabla A(S(x), \widetilde{y})\|^{2}\right] 
    + \mathbb{E}_{x, \widetilde{y}}\left[\|\widetilde{y}-S(x) \circ A(S(x), \widetilde{y})\|_{1}\right]
\end{equation}
Here, $x$ refers to the input image, while $\widetilde{y}$ represents its corresponding adjacent-slice image.

When dealing with unpaired data, the adjacency-guided loss is not calculated. So the loss function for unpaired data can be defined as follows:
\begin{equation}
l^{G} = \lambda_{1} \times (l^{G}_{adv} + l^{G}_{cls} + l_{cyc} + l_{idt})+\lambda_{2} \times(
l^{G}_{\eta adv} + l^{G}_{\eta cls})
\end{equation}

As for discriminator, the loss function remains consistent across both paired data and unpaired data. The discriminator's loss function consists of five components: adversarial losses $l^{D}_{adv1}$ and $l^{D}_{adv2}$, classification loss $l^{D}_{cls}$, auxiliary losses $l^{D}_{\eta adv}$ and $l^{D}_{\eta cls}$, and pre-training discriminator loss $l^{D}_{frz}$:
\begin{equation}
l^{D} =  \lambda_{1} \times (l^{D}_{adv} + l^{D}_{cls}) +  \lambda_{2} \times (l^{D}_{\eta adv} + l^{D}_{\eta cls}) + \lambda_{3} \times l^{D}_{frz}
\end{equation}

The details of the loss functions can be found in the supplementary materials.
\section{Experiments}

\subsection{Dataset}
To ensure the accuracy and reliability of our research data, we collaborated with experts from the pathology department of Peking University Third Hospital to conduct sample preparation and data collection. We obtained two sets of adjacent slices from each kidney tissue, with each set consisting of four slices stained with H\&E, Masson, PASM, and PAS, respectively. We ensured that the thickness of the slices was within the range of 1-2 $\mu m$. All tissue slices were extracted from pre-existing specimens, which were subjected to meticulous de-identification of any patient-related information. Therefore, this work does not impede conventional nursing practices or sample collection procedures. After acquiring the specimens, the slices were scanned using Jiangfeng Bio's KF-pro-005 whole slice scanner equipped with a ×40 objective lens. 

After the processing of the adaptive pairing module, we have successfully created a comprehensive renal histological dataset obtained from serial slices stained with H\&E, PAS, PASM, and Masson. The dataset contains 188 whole slide images from 22 patients with pathologists' WSI-level diagnoses, and 32,413 pairs of patch-level aligned glomeruli. In short, the glomerulus-aligned dataset fills the gap of high-quality, open-source multiple staining datasets in the field of renal histology. It will facilitate the evaluation and further development of virtual staining techniques.

\subsection{Implementation Details}

Our experiments were implemented on a device with an Intel(R) Xeon(R) Gold 5218 CPU, and one NVidia Telsa V100 GPU. We used the PyTorch framework to implement our algorithm. In the staining transfer process, we trained our model for 600,000 iterations.Following UMDST's\cite{lin2022unpaired} settings, we employed Adam optimizer with a learning rate of 1e-4 and linearly decaying it at 150,000 iterations. Both the training and testing batch sizes were set to 1.

\subsection{Results}
\begin{table}
\centering
{\caption{Comparison of different methods.}\label{table1}}
\begin{tabular}{lcccccc}
\hline
\rule{0pt}{12pt}
&\multicolumn{2}{c}{H\&E2MASS}&\multicolumn{2}{c}{H\&E2PASM}&\multicolumn{2}{c}{H\&E2PAS}\\
Methods&DISTS$\downarrow$&DBCNN$\uparrow$&DISTS$\downarrow$&DBCNN$\uparrow$&DISTS$\downarrow$&DBCNN$\uparrow$
\\
\hline
\\[-6pt]
FUNIT\cite{liu2019few}  & 0.2452 & 56.52 & 0.2496 & 54.01 & 0.2259 & 53.17   \\
MUNIT\cite{huang2018multimodal}  & 0.3207 & 29.76 & 0.3037 & 32.67 & 0.2764 & 36.02 \\
UGATIT\cite{kim2019u}   & 0.2374 & 57.18 & 0.2467 & 55.64 & 0.2166 & 51.91 \\
UMDST\cite{lin2022unpaired}  & 0.2064 & 53.44 & 0.2770 & 54.02 & 0.2162  & 51.90 \\
Ours(w/o Adj) & 0.2143 & 58.75 & 0.2468 & 60.61 & 0.1917  & 55.74 \\
Ours (w Adj) & \textbf{0.2020} & \textbf{64.49} & \textbf{0.2453} & \textbf{66.23} & \textbf{0.1909} & \textbf{59.85}\\
\hline
\\[-6pt]
\end{tabular}
\end{table}

Here, we present the stain transfer results of our proposed method and baselines. Our method outperforms all baselines in terms of both quantitative evaluation and visual effects. The evaluated baselines include UGATIT and UMDST, which have previously demonstrated effective stain transfer\cite{lin2022unpaired}, as well as the structurally simpler MUNIT\cite{huang2018multimodal} and FUNIT\cite{liu2019few} methods. Among these approaches, UGATIT and MUNIT require separate models for each specific stain transfer. In contrast, our approach, UMDST, and FUNIT achieve multi-domain staining transfer using a single model.

Table 1 presents the quantitative evaluation results of different methods. We use two metrics, DISTS\cite{ding2020image} and DBCNN\cite{zhang2018blind}, to assess the quality of stain transfer results. Both DISTS and DBCNN are effective indicators for evaluating image quality. The results in Table 1 indicate that our approach produces stained images of higher quality, with greater similarity to real adjacent-slice images and higher structural and textural similarity when compared to other methods for all three special staining transfer task.

Figure 3 illustrates the visual results of stain transfer using both AGMDT and baselines. The images generated by AGMDT exhibited correct anatomical structures and superior color mapping performance for all three stain transfer tasks. While MUNIT and FUNIT learned good style information, they failed to preserve anatomical structure accurately. UGATIT, UMDST, and our proposed method were able to retain the anatomical structure effectively. However, our method produced more realistic color mapping results that closely resembled adjacent-slice images. This improved realism is attributed to the effective incorporation of adjacent-slice information constraints in our framework.

\begin{figure*}[htbp]
\centerline{\includegraphics[width = 0.9\textwidth]{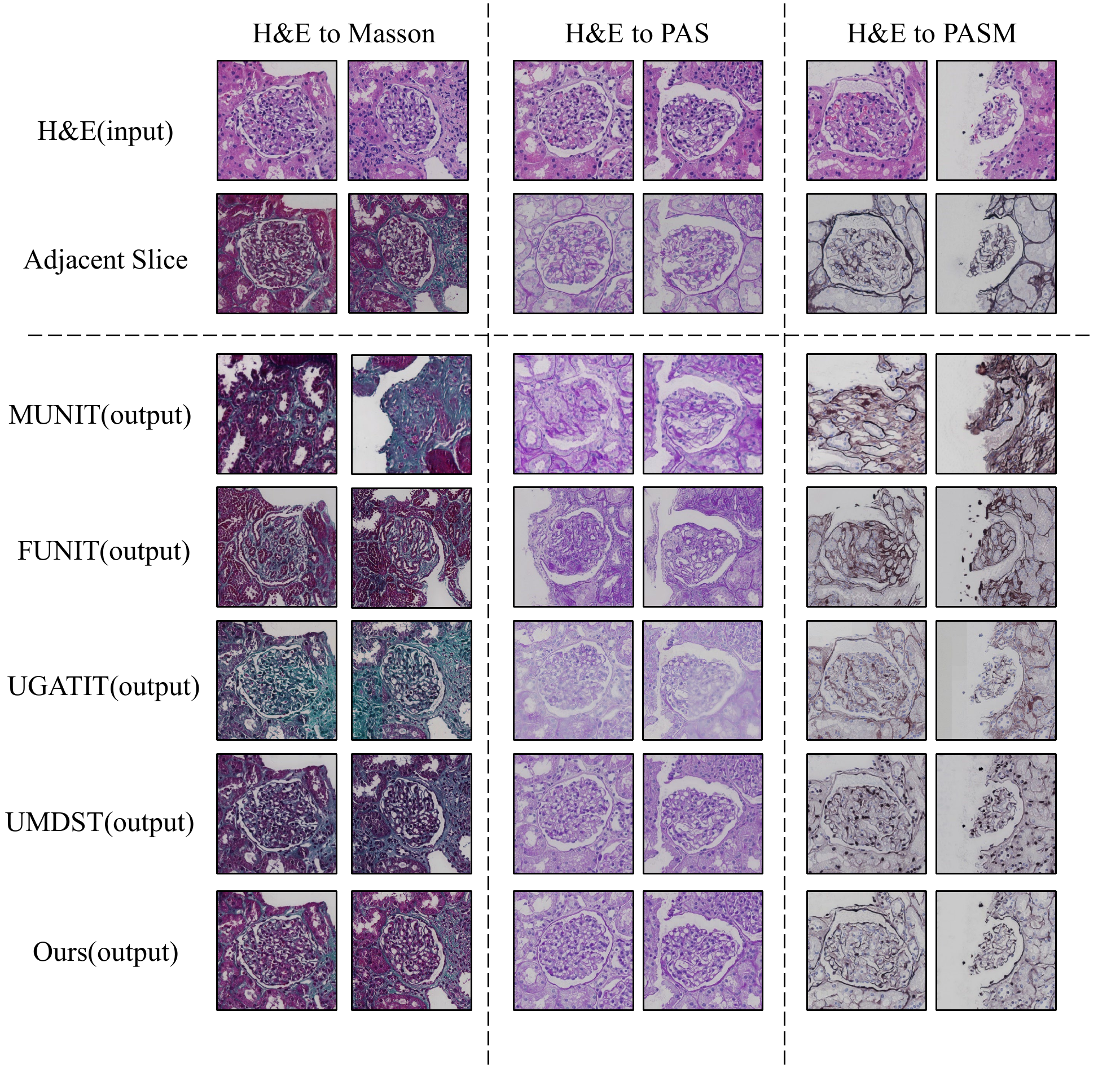}}
\caption{The virtual staining results of AGMDT and baseline methods.The first row shows the input H\&E image, and the second row represents the adjacent slice reference of the corresponding virtual staining target domain. Lines 3-7 show the virtual staining results of each method.} \label{Result01}
\end{figure*}

\subsection{Ablation Study}
We conducted an experiment to evaluate the impact of adjacent-slice information by comparing our model to a control model that excluded the adjacent supervision component while maintaining identical settings. As demonstrated in Figure 4, virtual staining results were generated for Masson, PASM, and PAS staining using both models - with and without the adjacent supervision component. For Masson virtual staining, the red rectangle in the figure illustrates that our model successfully learned the color mapping relationship between H\&E and Masson by incorporating adjacent-slice information constraints, while the model without these constraints failed to accurately map the glomerular internal tissue structure to red, instead displaying it as green. The adjacent supervision component also produced clearer and more realistic basement membrane structures for virtual PASM staining. Similar improvements can also be observed for PAS staining in terms of increased clarity and accuracy of basement membrane structure representation after the integration of adjacent-slice information constraints.

\section{Conclusion}
In this paper, we introduce an adjacency-guided multi-domain transfer framework to virtually transfer renal histology images into special staining types. The framework is a uniform one to generate multiple staining effects which includes a patch-level multi-domain adaptive paring module based on bipartite graph matching, and an adjacency-guided encoder with adjacency supervision. This is the first multi-modal staining transfer method to incorporate adjacent-slice information. 
A full-stack renal histological dataset based on adjacent slices of the four typical staining is also created, with pathologists' WSI-level diagnoses and 32,413 pairs of patch-level aligned glomeruli.
Experimental results show that this method achieves the best virtual staining effect compared to state-of-the-art methods in terms of both quantitative measurements and visual details for clinical diagnosis. AGMDT framework can also be extended to virtual staining for other types such as IHC, given the adjacent-slice information is included in the training dataset.

\section*{ACKNOWLEDGEMENTS}
We thank Qiuchuan Liang for doing some data processing work.
\begin{figure}[htbp]
\centerline{\includegraphics[width = 0.8\textwidth]{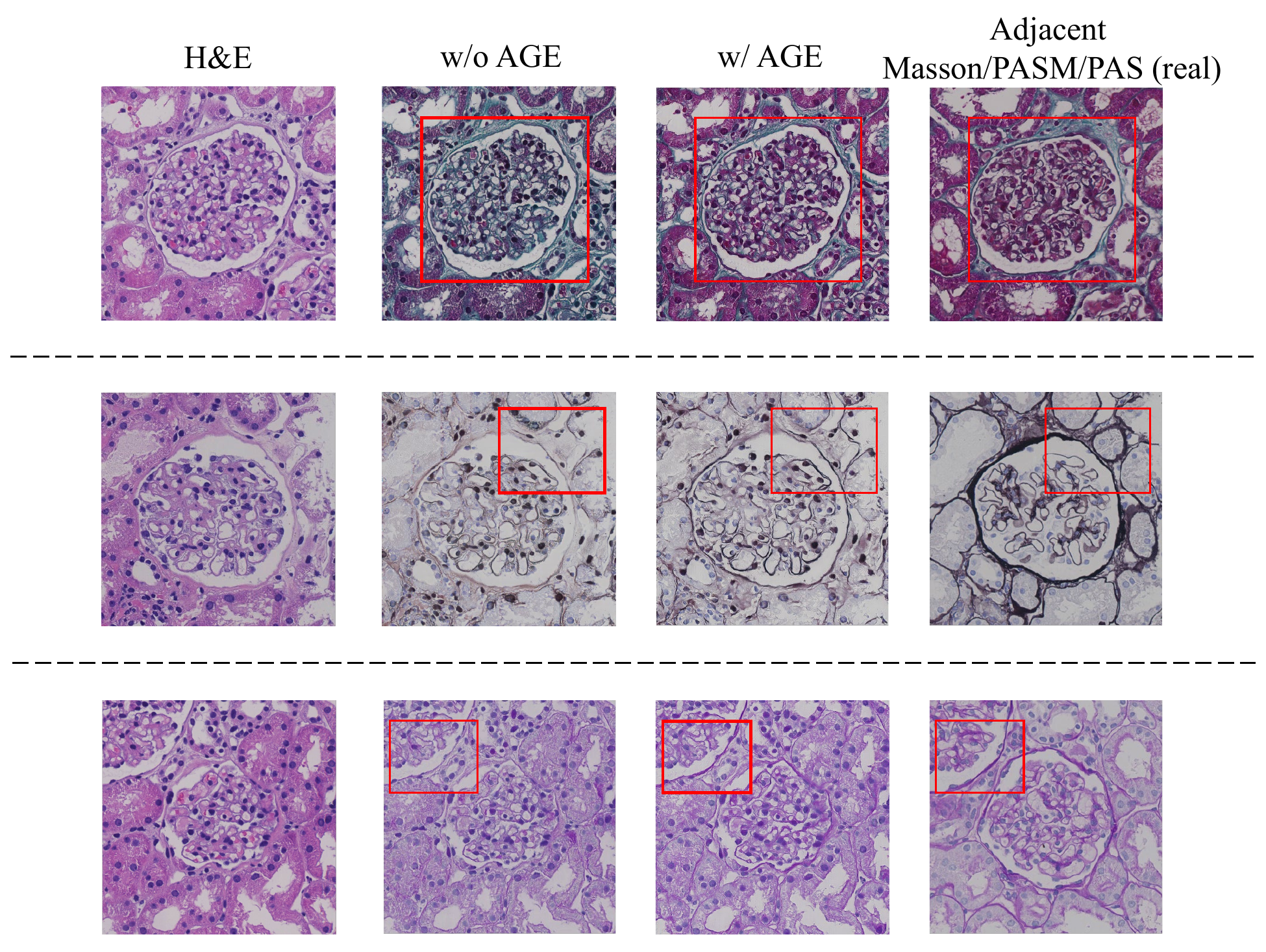}}
\caption{The virtual staining results of our model with or without the AGE module.} \label{Ablation}
\end{figure}

\bibliography{egbib}
\end{document}